\RequirePackage{lineno}
\documentclass[a4paper,10pt,oneside]{cpc-hepnp}
\usepackage{multicol}
\usepackage{CJK,upgreek,fancyhdr}
\usepackage{booktabs}
\usepackage{mathrsfs}
\usepackage{bbm}
\usepackage{amscd}
\usepackage[tbtags]{amsmath}
\usepackage{amssymb,bm,mathrsfs,bbm,amscd}
\usepackage{lastpage}
\usepackage{upgreek,fancyhdr}
\usepackage{color}
\usepackage{geometry}
\usepackage{xspace}
\usepackage{multirow}
\usepackage{hhline}
\usepackage{amssymb}
\usepackage{lineno}
\usepackage{overpic}
\usepackage{graphicx}
\usepackage{float}
\usepackage{dcolumn}
\usepackage{bm}
\usepackage{rotating}
\usepackage{epsfig,graphics,subfigure,psfrag}
\usepackage[colorlinks=true,linkcolor=blue,anchorcolor=black,citecolor=blue]{hyperref}
\let\oldequation\equation
\let\oldendequation\endequation

\renewenvironment{equation}
  {\linenomathNonumbers\oldequation}
  {\oldendequation\endlinenomath}

\begin{document}
\begin{CJK*}{UTF8}{gkai}
\fancyhead[co]{\footnotesize \boldmath Observation of $e^+e^- \to p p \bar{p} \bar{n} \pi^{-} + c.c.$}

\footnotetext[0]{Received xxxx june xxxx}

\title{\boldmath Observation of $e^+e^- \to p p \bar{p} \bar{n} \pi^{-} + c.c.$
\thanks{
This work is supported in part by National Key R\&D Program of China under Contracts Nos. Supported
in part by National Key R\&D Program of China under Contracts
Nos. 2020YFA0406300, 2020YFA0406400; National Natural Science
Foundation of China (NSFC) under Contracts Nos. 11975118, 11625523, 11635010,
11735014, 11822506, 11835012, 11935015, 11935016, 11935018,
11961141012, 12022510, 12025502, 12035009, 12035013,
12061131003, 12075252, 12192260, 12192261, 12192262, 12192263, 12192264, 12192265; the Natural Science Foundation of Hunan Province of China under Contract No. 2019JJ30019;
the Science and Technology Innovation Program of Hunan Province under Contract No. 2020RC3054;
the Chinese Academy of Sciences (CAS)
Large-Scale Scientific Facility Program; Joint Large-Scale Scientific
Facility Funds of the NSFC and CAS under Contracts Nos. U1732263,
U1832207; CAS Key Research Program of Frontier Sciences under Contract
No. QYZDJ-SSW-SLH040; 100 Talents Program of CAS; INPAC and Shanghai
Key Laboratory for Particle Physics and Cosmology; ERC under Contract
No. 758462; European Union Horizon 2020 research and innovation
programme under Contract No. Marie Sklodowska-Curie grant agreement No. 894790; German Research Foundation DFG under Contracts Nos. 443159800,
Collaborative Research Center CRC 1044, FOR 2359, GRK 2149; Istituto Nazionale di Fisica Nucleare, Italy; Ministry of Development of Turkey under Contract No. DPT2006K-120470; National Science and Technology fund; Olle Engkvist Foundation under Contract No. 200-0605; STFC (United Kingdom); The Knut and Alice Wallenberg Foundation (Sweden) under Contract No. 2016.0157; The Royal Society, UK under Contracts Nos. DH140054, DH160214; The Swedish Research Council; U. S. Department of Energy under Contracts Nos. DE-FG02-05ER41374, DE-SC-0012069.
}
\vspace{-0.5in}
}
\maketitle

\begin{small}
\begin{center}
\begin{small}
\begin{center}
M.~Ablikim$^{1}$, M.~N.~Achasov$^{11,b}$, P.~Adlarson$^{70}$, M.~Albrecht$^{4}$, R.~Aliberti$^{31}$, A.~Amoroso$^{69A,69C}$, M.~R.~An$^{35}$, Q.~An$^{53,66}$, X.~H.~Bai$^{61}$, Y.~Bai$^{52}$
% ...
\\
\vspace{0.2cm}
(BESIII Collaboration)\\
\vspace{0.2cm} {\it
$^{1}$ Institute of High Energy Physics, Beijing 100049, People's Republic of China\\
% ...
$^{4}$ Bochum Ruhr-University, D-44780 Bochum, Germany\\
% ...
$^{11}$ G.I. Budker Institute of Nuclear Physics SB RAS (BINP), Novosibirsk 630090, Russia\\
% ...
$^{31}$ Johannes Gutenberg University of Mainz, Johann-Joachim-Becher-Weg 45, D-55099 Mainz, Germany\\
% ...
$^{35}$ Liaoning Normal University, Dalian 116029, People's Republic of China\\
% ...
$^{52}$ Southeast University, Nanjing 211100, People's Republic of China\\
$^{53}$ State Key Laboratory of Particle Detection and Electronics, Beijing 100049, Hefei 230026, People's Republic of China\\
% ...
$^{61}$ University of Jinan, Jinan 250022, People's Republic of China\\
% ...
$^{66}$ University of Science and Technology of China, Hefei 230026, People's Republic of China\\
% ...
$^{69}$ (A)University of Turin, I-10125, Turin, Italy; (B)University of Eastern Piedmont, I-15121, Alessandria, Italy; (C)INFN, I-10125, Turin, Italy\\
$^{70}$ Uppsala University, Box 516, SE-75120 Uppsala, Sweden\\
% ...
$^{b}$ Also at the Novosibirsk State University, Novosibirsk, 630090, Russia\\
% ...
}\end{center}

\vspace{0.4cm}
\end{small}
\end{center}
\vspace{-0.2cm}
\end{small}

\begin{abstract}
Using data taken at 29 center-of-mass energies between 4.16 and 4.70 GeV with the BESIII detector at the Beijing Electron Positron Collider corresponding to a total integrated luminosity of approximately 18.8 $\rm fb^{-1}$,
the process $e^+e^- \to p p \bar{p} \bar{n} \pi^{-} + c.c.$
is observed for the first time with a statistical significance of $11.5\sigma$.
The average Born cross sections in the energy ranges of (4.160, 4.380) GeV, (4.400, 4.600) GeV and (4.610, 4.700) GeV are
measured to be $(21.5\pm5.7\pm1.2)$ fb, $(46.3\pm10.6\pm2.5)$ fb and $(59.0\pm9.4\pm3.2)$ fb, respectively, where the first uncertainties are statistical and the second are systematic.
The line shapes of the $\bar{p}\bar{n}$ and $pp\pi^-$ invariant mass spectra are consistent with phase space distributions, indicating that no hexaquark or di-baryon state is observed.
\end{abstract}


\begin{multicols}{2}
\section{INTRODUCTION}

One of the most fundamental questions in hadron physics is related to the mechanism of color confinement in
Quantum Chromodynamics (QCD).
Color-singlet hadronic configurations of quarks and gluons can
form bound states or resonances. Besides the well-known combinations of
$q\bar{q}$ for mesons and
$qqq$ for baryons, other combinations, such as $gq\bar{q}$ for hybrid states~\cite{hybrid}, multi-gluons for glueball states~\cite{glueball},
$q\bar{q}q\bar{q}$ for tetraquark states~\cite{pipijpsi}, $qqqq\bar{q}$ for pentaquark states~\cite{jpsip} and hexaquark states $(qqqqqq)$, are also allowed by QCD.
Di-baryon and hexaquark states have been searched for in a range of nucleon-nucleon scattering reactions.
Recently, an isoscalar resonant structure was observed in the isoscalar two-pion fusion process $pn \to d\pi^0\pi^0$~\cite{dp0p0}
by the WASA Collaboration and was later confirmed in the other two-pion fusion processes $pn\to d\pi^+\pi^-$~\cite{dpppm} and $pp\to d\pi^+\pi^0$~\cite{dppp0},
and the two-pion non-fusion process $pn\to pp\pi^0\pi^-$~\cite{ppp0pm} and $pn\to pn\pi^0\pi^0$~\cite{pnp0p0}.
This state was denoted by $d^*(2380)$ following the convention used
for nucleon excitations.
These observations indicate the possibility of the existence of hexaquark and di-baryon configurations.
In 2021, the BESIII Collaboration reported the search for hexaquark and di-baryon states
in examining the invariant mass spectra of two baryons in the process $e^+e^- \to 2(p\bar{p})$~\cite{pppp},
and no significant signal was observed.

Analyzing data sets corresponding to a total integrated luminosity of approximately $18.8$ $\rm fb^{-1}$ taken at
center-of-mass energies $\sqrt s$ between $4.16$ and $4.70$ GeV with
the BESIII detector, we present in this paper
the first measurement of the cross section of the process $e^+e^- \to
pp\bar{p}\bar{n}\pi^{-} + c.c.$.  We search for the $d^*(2380)$ and other possible hexaquark or di-baryon states
with the data samples with energies above $4.60$ GeV, where the $\bar{p}\bar{n}$ system with a mass around $2.4$ GeV for $d^*(2380)$ is kinematically accessible.
The mass range of the $\bar{p}\bar{n}$ system around 2.4 GeV/$c^2$, in which the $d^*(2380)$ might contribute, is covered by the data samples with energies above $4.60$ GeV.
Throughout this paper, charge conjugation is always implied unless explicitly stated, and in discussing systematic uncertainties.

\section{THE BESIII DETECTOR AND DATA SAMPLES}

The BESIII detector~\cite{detector} records symmetric $e^+e^-$ collisions provided by the
BEPCII storage ring~\cite{BEPCII}, which operates
in the center-of-mass energy range from $2.0$ to $4.95$ GeV. BESIII has collected large data samples in this energy
region~\cite{samples}. The cylindrical core of the BESIII detector covers $93\%$ of the full solid
angle and consists of a helium-based multilayer drift chamber (MDC), a plastic scintillator
time-of-flight system (TOF), and a CsI(Tl) electromagnetic calorimeter (EMC), which are
all enclosed in a superconducting solenoidal magnet providing a $1.0$~T
magnetic field. The solenoid is supported by an octagonal flux-return yoke with resistive
plate counter muon identification modules interleaved with steel. The charged-particle momentum
resolution at $1~{\rm GeV}/c$ is $0.5$\%, and the specific energy loss
(d$E$/d$x$) resolution is $6\%$ for electrons from
Bhabha scattering. The EMC measures photon energies with a resolution of $2.5\%$ ($5\%$) at
$1$ GeV in the barrel (end cap) region. The time resolution in the TOF barrel region is $68$ ps,
while that in the end cap region is $110$ ps. The end cap TOF system was upgraded in $2015$
using multi-gap resistive plate chamber technology, providing a time resolution of $60$ ps~\cite{tof}.

The data sets were collected at $29$ center-of-mass energies between $4.16$ and $4.70$ GeV.
The nominal energies of the data sets from $4.16$ to $4.60$ GeV are measured by the di-muon process
$e^+e^- \to (\gamma_{\rm ISR/FSR})\mu^{+}\mu^{-}$~\cite{mumu1, mumu2}, where
the subscript ISR/FSR stands for the initial-state or final-state
radiation process, respectively. The data sets from $4.61$ to $4.70$ GeV are calibrated by the process $e^+e^- \to \Lambda_{\rm c}^{+}\bar{\Lambda}_{\rm c}^{-}$~\cite{LcLc}.
 The integrated luminosity $\mathcal{L}_{\rm int}$ is determined using large-angle
Bhabha scattering events~\cite{LcLc,rlum}.  The total integrated luminosity of all data sets is approximately $18.8$ $\rm fb^{-1}$.

The response of the BESIII detector is modeled with Monte Carlo (MC) simulations using the
software framework {\sc boost}~\cite{boost} based on {\sc geant4}~\cite{geant4}, which
includes the geometry and material description of the BESIII detector, the detector
response and digitization models, as well as a database that keeps track of the running
conditions and the detector performance. Large MC simulated event samples are used to optimize the selection
criteria, evaluate the signal efficiency, and estimate background contributions.

Inclusive MC simulation samples are generated at different center-of-mass energies to study potential background reactions.
These samples consist of open charm processes, the ISR production of
vector charmonium and charmonium-like states, and the continuum processes incorporated in
{\sc kkmc}~\cite{kkmc}. The known decay modes are modeled with {\sc evtgen}~\cite{besevtgen}
using branching fractions taken from the Particle Data Group (PDG)~\cite{pdg2013}, and the remaining
unknown decays of the charmonium states are simulated with {\sc lundcharm}~\cite{lundcharm}. Final-state radiation from
charged final-state particles is incorporated with {\sc photos}~\cite{photos}.
The signal MC simulation sample of $e^+e^- \to p p \bar{p} \bar{n} \pi^{-} $ at each energy point is generated with the events being uniformly distributed in phase space.

\section{DATA ANALYSIS}

Events with two positive and two negative charged tracks are selected.
For each charged track, the polar angle in the MDC with respect to the $z$ direction must satisfy $|\!\cos\theta|<0.93$.
All charged tracks are required to
originate from the interaction region, defined as $R_{xy}<1$~cm and $|V_{z}|<10$~cm, where $R_{xy}$ and $|V_{z}|$
are the distances from the point of closest approach of the tracks to the interaction point
in the $x-y$ plane and in the $z$ direction, respectively.
The combined d$E$/d$x$ and TOF information are used to calculate
particle identification (PID) confidence levels for the pion, kaon, and proton hypotheses.
Each track is assigned as the particle hypothesis with the highest confidence
level. The final state in the $e^+e^- \to p p \bar{p} \bar{n} \pi^{-}$ process is reconstructed with three (anti-)protons
and one $\pi^{-}$.

Since the neutron can not be well reconstructed with the BESIII
detector, the signal process is determined via the recoiling mass of the reconstructed charged particles,
defined as
\begin{equation} \label{eq:Mn}
M_{\rm rec}c^2= \sqrt{(E_{e^{+}e^{-}} - E_{pp\bar{p}\pi^{-}})^{2} - (\vec{P}_{e^{+}e^{-}} - \vec{P}_{pp\bar{p}\pi^{-}})^{2} \cdot c^2},
\end{equation}
where $E_{e^{+}e^{-}}$ and $\vec{P}_{e^{+}e^{-}}$ are the center-of-mass energy and the momentum of the $e^{+}e^{-}$ system, respectively;
$E_{pp\bar{p}\pi^{-}}$ and $\vec{P}_{pp\bar{p}\pi^{-}}$ are the total reconstructed energy and total momentum of the $pp\bar{p}\pi^{-}$ system, respectively.
Events with $M_{\rm rec}$ greater than $0.8$ GeV/$c^2$ are kept for further analysis.

Studies based on the inclusive MC simulation samples~\cite{topoana} show that no peaking background events survive the selection criteria.
To further suppress background events, two additional selection criteria are imposed on the accepted candidate events.
First, the invariant mass $M_{p\pi^-}$ of the reconstructed $p\pi^-$ system is required to
be outside the $\Lambda$ signal region, $i.e.$ $|M_{p\pi^-}-1.115|>0.010$ GeV/$c^2$,
to remove the possible background associated with $\Lambda$ decays. Here, 1.115~GeV/$c^2$ is the known $\Lambda$ mass~\cite{pdg}, and 0.010~GeV/$c^2$ corresponds to about three times the mass resolution.
Second, the invariant mass of $pp\bar{p}$ ($M_{pp\bar{p}}$) must be less than $3.6$ GeV/$c^2$ due to the remaining neutron and pion in the event.

\begin{figure*}[tbh]
	\centering
	\includegraphics[width=0.45\textwidth]{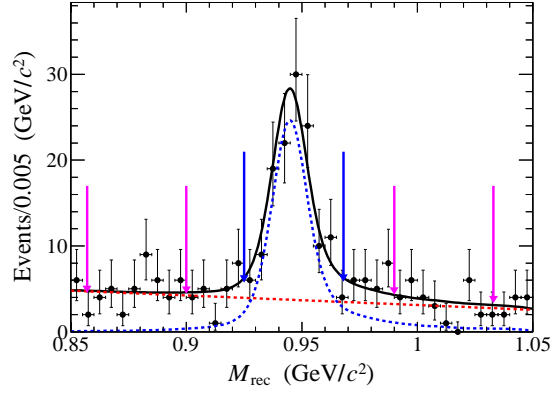}
    \caption{Distribution of the recoiling mass $M_{\rm rec}$ of the candidate events for the reaction $e^+e^-\to p p \bar{p} \bar{n} \pi^{-}$ with the fit results overlaid. The dots with error bars are from the combined data sets,
    the black curve shows the total fit result, and the dashed blue (red) curve represents
     the signal (background) shape.
     The pair of blue arrows marks the neutron signal region, whereas the neutron sideband regions are visualized by the two pairs of pink arrows.}
	\label{fig:Mnfit}
\end{figure*}

The $M_{\rm rec}$ distribution of the accepted candidates after the above selection criteria from the combined data sets is displayed in Fig.~\ref{fig:Mnfit},
where a significant neutron signal is observed.
The signal yield is determined by a maximum likelihood fit to this distribution. In the fit, the signal is
represented by the luminosity weighted MC-simulated shape convolved with a Gaussian function and the remaining background is described by a linear function.
From the fit, the signal yield is determined to be $123\pm14$.
The statistical significance of the signal is determined to be $11.5\sigma$, which is evaluated
as $\sqrt{-2\ln(\mathcal{L}_0/\mathcal{L}_{\rm max})}$, where $\mathcal{L}_{\rm max}$ is
the maximum likelihood of the nominal fit and $\mathcal{L}_0$ is the likelihood of the fit
without involving the signal component. The change of the degree of freedom is $1$.
The neutron signal region is defined as $M_{\rm rec}\in(0.925,0.968)$ GeV/$c^2$ and
the corresponding sideband regions are defined as $M_{\rm rec}\in$ (0.857,0.900) $\cup$  (0.990,1.033) GeV/$c^2$.

\begin{figure*}[!tbh]
	\centering
	\includegraphics[width=0.45\textwidth]{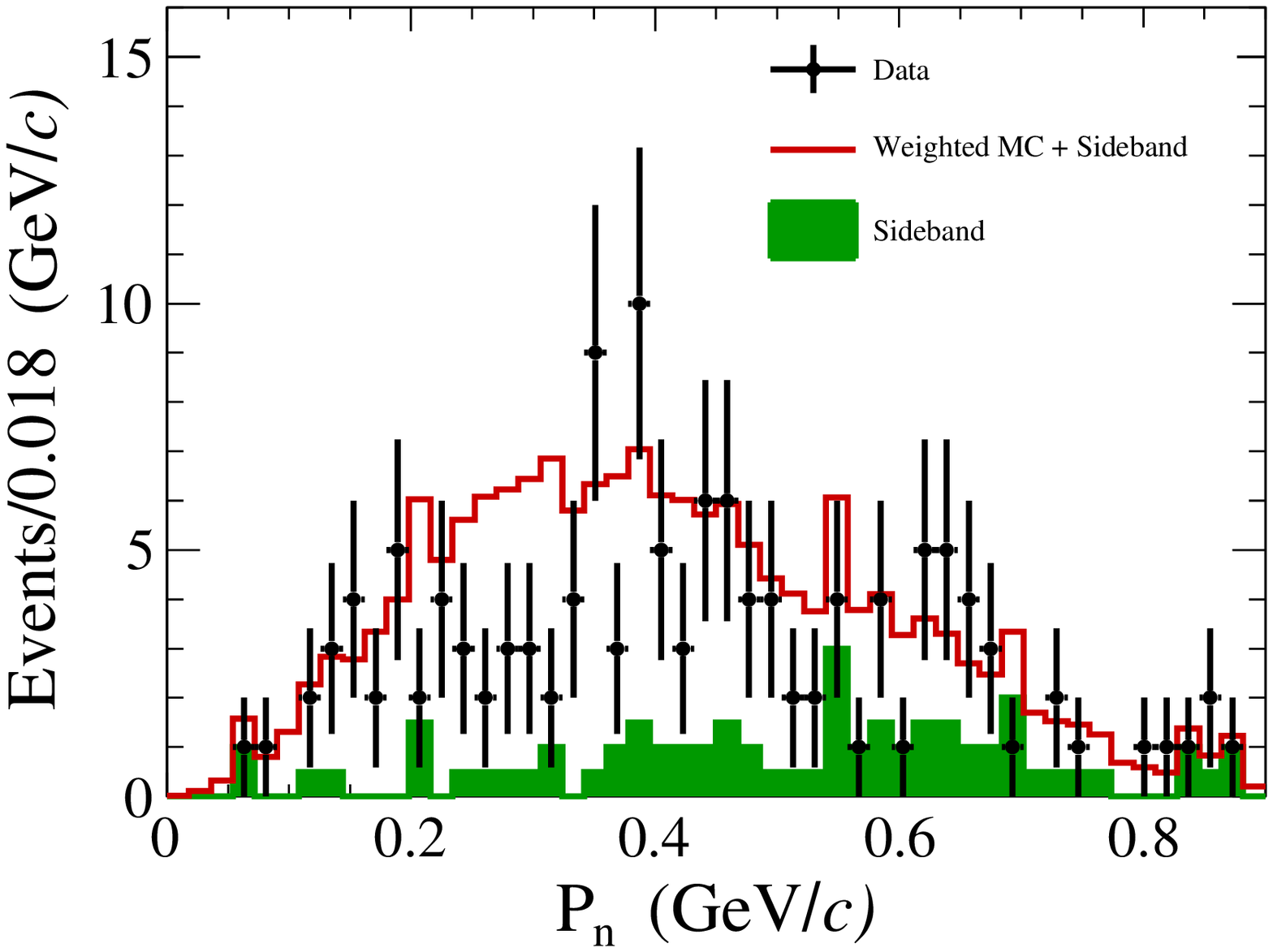}
    \includegraphics[width=0.45\textwidth]{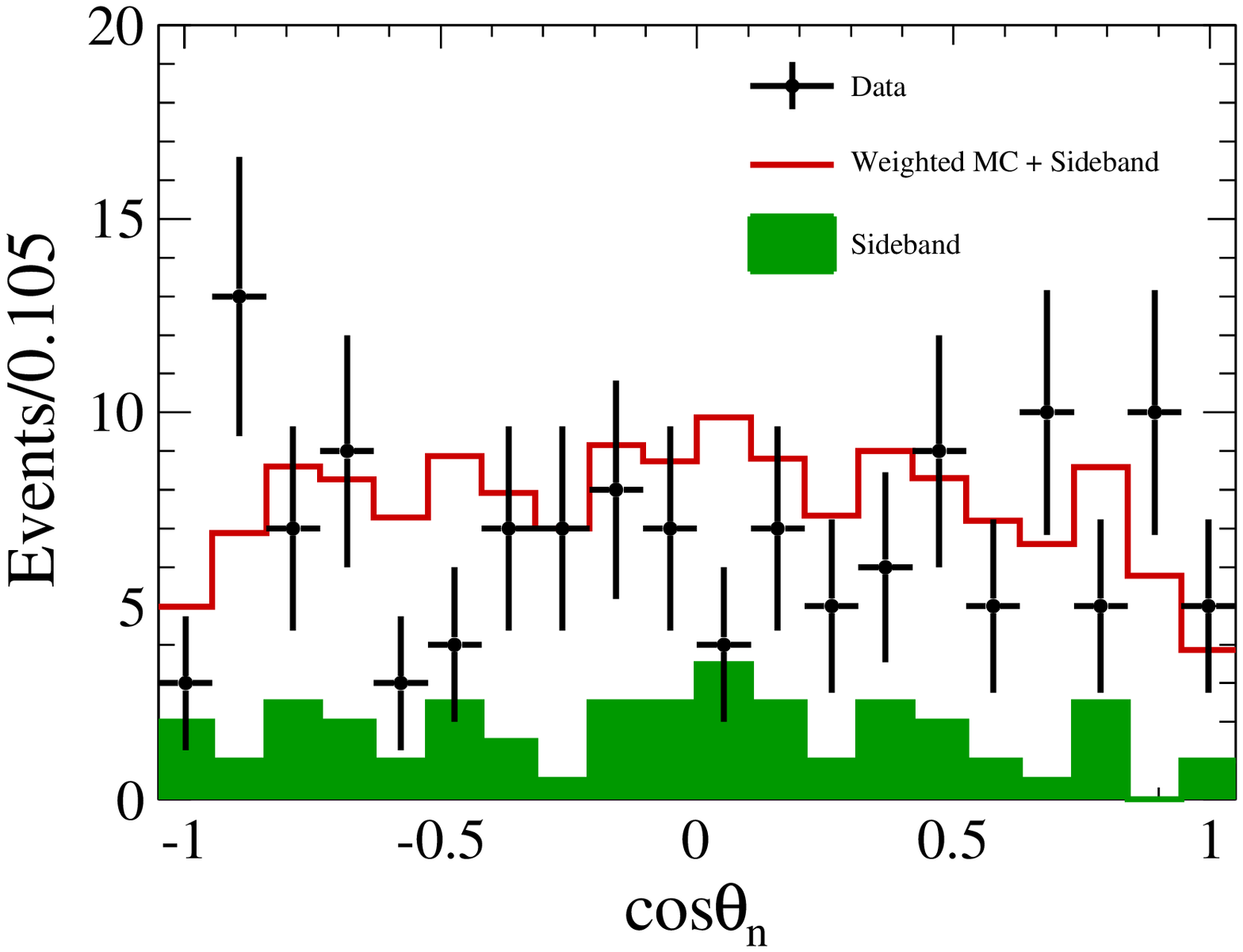}
	\caption{\small Momentum (left) and polar angular distributions (right) of the neutrons of the candidate events for the reaction $e^+e^-\to p p \bar{p} \bar{n} \pi^{-}$.
The dots with error bars represent the combined data sets. The green histograms represent the sideband events.
The red histograms represent the weighted signal MC simulation events plus the normalized neutron sideband regions in data.}
	\label{fig:Momentum}
\end{figure*}

Figure~\ref{fig:Momentum} shows the comparisons of the momentum and polar angle distributions of the neutron
of the accepted candidate events between data and signal MC simulation samples,
where the data distribution is from the combined data sets
and the MC simulation distribution has been weighted by the signal yields in data.
The invariant mass of any two or three particles,
the momentum and $\cos\theta$ distributions of the other final state particles have also been examined.
The agreement between data and MC simulation allows to determine the detection efficiency with the signal MC simulation events
generated uniformly distributed in the five-body phase space.

\begin{figure*}[!tbh]
	\centering
    \includegraphics[width=0.45\textwidth]{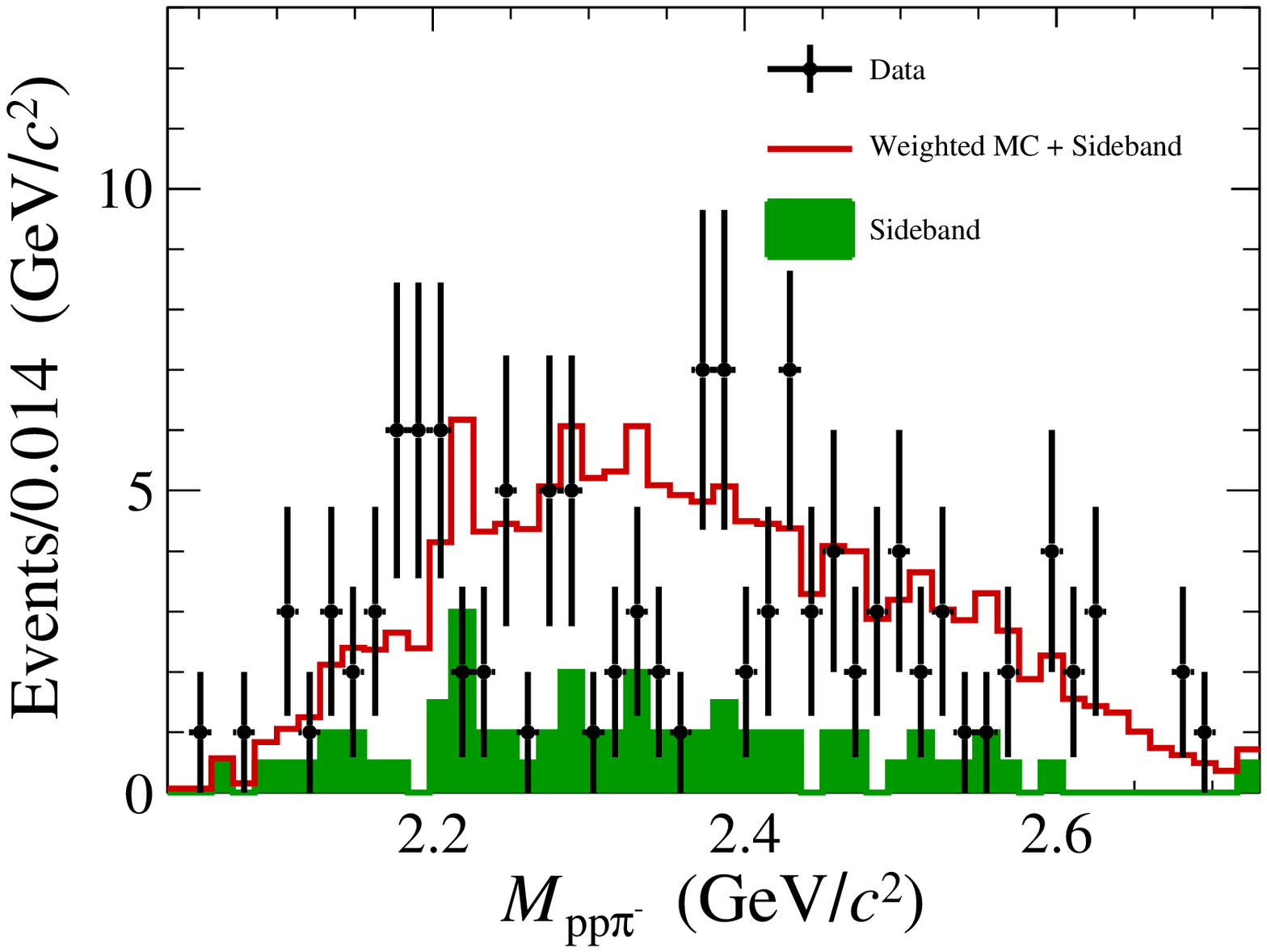}
	\includegraphics[width=0.45\textwidth]{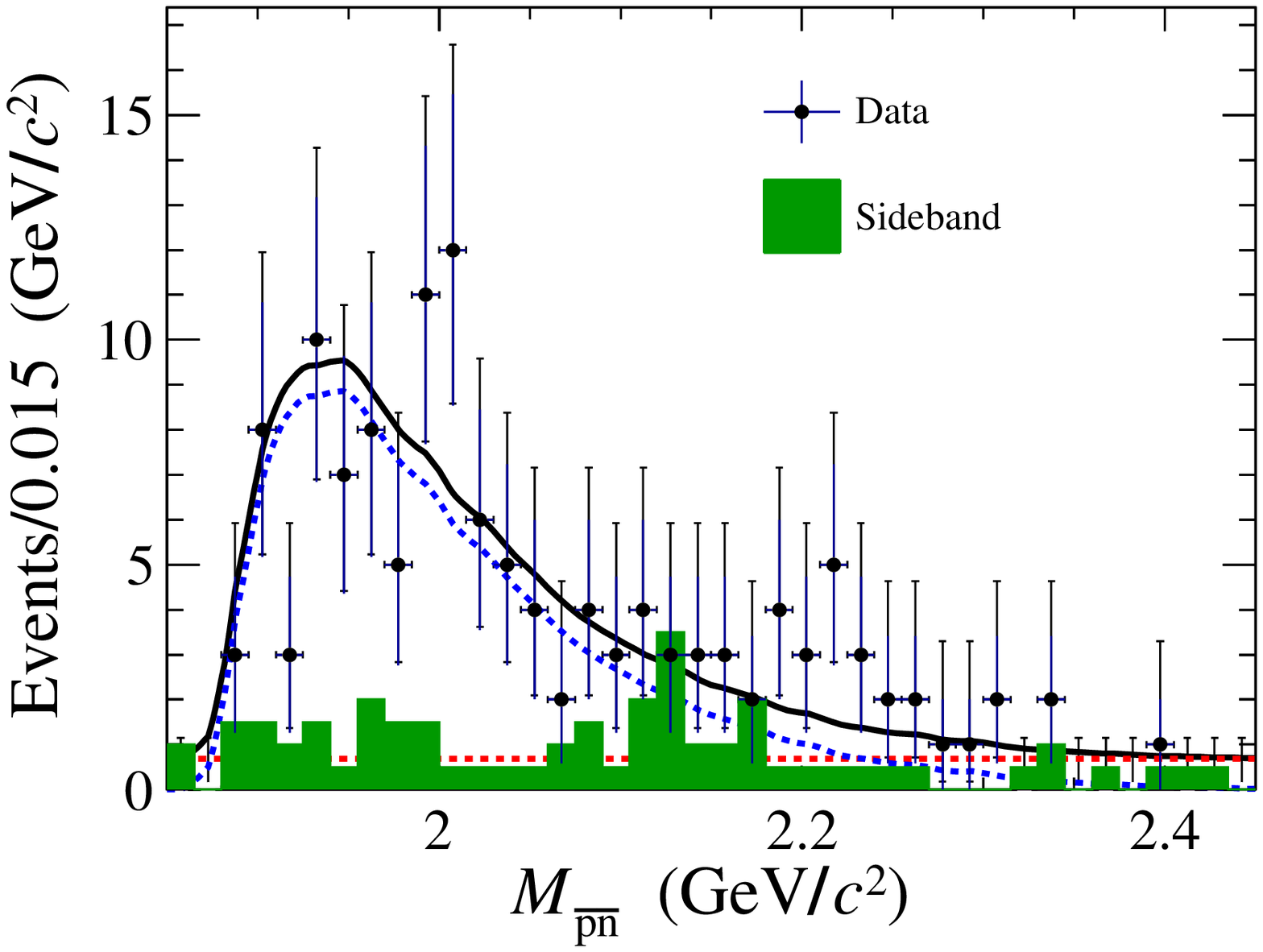}
	\caption{\small The $pp\pi^{-}$ (left) and $\bar{p}\bar{n}$ (right) invariant mass spectra of the candidate events for $e^+e^-\to p p \bar{p} \bar{n} \pi^{-}$.
The dots with error bars represent the combined data sets.
The green histograms are the normalized neutron sideband events in data.
The red histograms represent the weighted signal MC simulation events plus the normalized neutron sideband events in data.
The black solid curve shows the total fit result,
the blue dashed curve is the signal shape derived from the signal MC simulation sample,
and the red dashed line is the linear background shape.}
	\label{fig:Mpn}
\end{figure*}

To search for hexaquark and di-baryon states, the $\bar{p}\bar{n}$ invariant mass spectrum is examined. Figure~\ref{fig:Mpn} shows the $pp\pi^{-}$ and $\bar{p}\bar{n}$ invariant mass spectra of the candidate events for the reaction $e^+e^-\to p p \bar{p} \bar{n} \pi^{-}$. In the fit to $M_{\bar{p}\bar{n}}$, the signal is
represented by the luminosity weighted phase space MC simulation shape
and the remaining combinatorial background is described by a linear
function.  The goodness-of-fit is $\chi^2/ndf= 2.10/2$. Here, $ndf$ is the
number of degrees of freedom. Compared to the phase space hypothesis, no obvious structure is observed.

\section{AVERAGE CROSS SECTIONS}
\label{sec:avgx}

\label{subsec:avgX}
In each data set, only a few events have been observed in the neutron signal region, with a statistical significance of less than $3\sigma$.
To obtain significant neutron signals the data sets are combined into three sub-samples in the energy ranges of (4.160, 4.380), (4.400, 4.600) and (4.610, 4.700) GeV
for further analysis.

The average observed cross section for $e^+e^- \to p p \bar{p} \bar{n} \pi^{-}$ is calculated by
\begin{equation} \label{eq:sigmaobsa}
\overline{\sigma}_j^{\rm obs} = \frac{N_j^{\rm sig}}{ {\Sigma_i {\epsilon_{ ji}}} \cdot {\mathcal{L}_{ ji}} } ,
\end{equation}
where $N_j^{\rm sig}$ is the number of signal events from the $j$-th combined data set,
$\mathcal{L}_{i}$ and $\epsilon_{i}$ are the integrated luminosity and efficiency of data set $i$, , respectively, $i$ represents the $i$-th
energy point in $j$-th sub-data set.
The detection efficiency is corrected by the PID and tracking efficiencies correction factors, ${f_{\rm PID}}$ and ${f_{\rm trk}}$, which are determined
to be $0.92$ and $0.98$ by weighting the differences between data and MC simulation efficiencies in different momentum ranges,
respectively. Inserting the numbers which are listed in Table~\ref{tab:sec} into Eq.~\ref{eq:sigmaobsa} yields the average observed cross sections $(19.4\pm5.1\pm1.0)$ fb, $(42.8\pm9.8\pm2.3)$ fb and $(54.2\pm8.6\pm2.9)$ fb for the three sub-data sets, respectively, where the first uncertainties are statistical and the second are systematic.

\begin{figure}[H]
	\centering
        \includegraphics[width=0.45\textwidth]{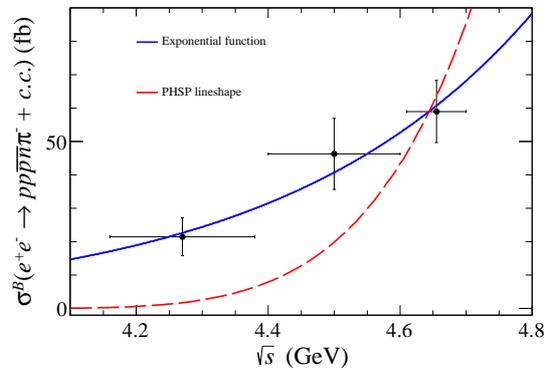}
	\caption{\small
Average Born cross sections for the process $e^+e^-\to p p \bar{p} \bar{n} \pi^{-} + c.c.$.
The dots with error bars are the measured values, the blue line shows the exponential function curve and the red-dashed line shows the five-body energy-dependent phase spase (PHSP) lineshape curve. }
	\label{fig:fitXsection}
\end{figure}

To measure the average Born cross section of $e^+e^- \to p p \bar{p} \bar{n} \pi^{-}$, a similar lineshape as that of $e^+e^- \to 2(p\bar{p})$~\cite{pppp} is
assumed to determine the ISR and vacuum polarization correction factors, $(1+\delta^{\gamma})$ and $\frac{1}{\vert 1-\Pi\vert^2}$, as they are similar reactions where one of the $\bar{p}$ has been exchanged by $\bar{n}\pi^-$.
The average Born cross section is calculated by
\begin{equation} \label{eq:sigmaborn}
\overline{\sigma}_j^{\rm Born} = \frac{\overline{\sigma}_j^{\rm obs}}{(1+\delta^{\gamma})_j \cdot \Big(\frac{1}{\vert 1-\Pi\vert^2}\Big)_j} .
\end{equation}
The obtained Born cross sections are then used as input in the generator and the cross section measurements are iterated with the updated detection efficiencies. This process is repeated
until the $(1+\delta^{\gamma}) \cdot \epsilon$ values become stable at all energies,
$i.e.$ the difference of $(1+\delta^{\gamma}) \cdot \epsilon$ between the last two iterations is less than $4\%$.
Figure~\ref{fig:fitXsection} shows the obtained average Born cross sections in the defined sub-samples.
The average Born observed cross sections are calculated with Eq.~\ref{eq:sigmaborn}, and the results are $(21.5\pm5.7\pm1.2)$ fb, $(46.3\pm10.6\pm2.5)$ fb and $(59.0\pm9.4\pm3.2)$ fb for the three sub-data sets, respectively, where the first uncertainties are statistical and the second are systematic.
Two different functions are used to compare the trend of the average Born cross section to a reaction where a similar behaviour is expected. The first one is a simple five-body energy-dependent phase space lineshape~\cite{pppp,fitting1} and the second one is an exponential function~\cite{pppp,exp2}, which are shown in Figure~\ref{fig:fitXsection}.
The exponential function is constructed as
\begin{equation} \label{eq:fitequ1}
\sigma^{\rm Born}(s) = \frac{1}{s} \times e^{-p_{\rm0}(\sqrt{s}-M_\text{th})} \times p_{\rm 1},
\end{equation}
where $p_{0}$ and $p_{1}$ are free parameters, $M_\text{th} = (3m_{p}+m_{n}+m_{\pi^-})$,
$m_p$, $m_n$, and $m_{\pi^-}$ are the known masses of $p$, $n$, and $\pi^-$ taken from the PDG~\cite{pdg}.
This is similar to the one used for the cross section lineshape of $e^+e^- \to 2(p\bar{p})$ in Ref.~\cite{pppp}, as they are similar reactions where one of the $\bar{p}$ has been exchanged by $\bar{n}\pi^-$.
However, it should be noted that the two functions in Figure~\ref{fig:fitXsection} are not fit results, but drawn with arbitrary scale factors for comparison since a qualitative fit is not possible due to the limited statistics.

The systematic uncertainties in the cross section measurements will be discussed in the next section.

\begin{table*}[htbp]
\begin{center}
\caption{The average observed cross sections for the reaction $e^+e^- \to p p \bar{p} \bar{n} \pi^{-} + c.c.$.
Summary of the number of signal events ($N_{\rm sig}$), integrated luminosity ($\mathcal{L}$), detection efficiency ($\epsilon$), radiative correction factors $(1+\delta^{\gamma})$, the average observed cross section ($\overline{\sigma}^{\rm obs}$) and average Born cross section ($\overline{\sigma}^{\rm Born}$) at different c.m. energies ($\sqrt{s}$). The uncertainties are  statistical only. } \label{tab:sec}
\begin{small}
\begin{tabular}{ccccccccc}\hline
& $\sqrt{s}$~(GeV)   &$\mathcal{L}$~(pb$^{-1}$)   & $\epsilon$ (\%)  &$(1+\delta^{\gamma})$  & $N_{\rm sig}$   & $\overline{\sigma}^{\rm obs}$ (fb)  & $\overline{\sigma}^{\rm Born}$ (fb) \\  \hline
&4.1574                   & 408.70                     & 4.2  & 0.8408 \\
&4.1780                   & 3189.0                     & 6.0  & 0.8388 \\
&4.1889                   & 526.70                     & 7.1  & 0.8430 \\
&4.1990                   & 526.00                     & 8.1  & 0.8456 \\
&4.2092                   & 517.10                     & 8.8  & 0.8485 \\
&4.2188                   & 514.60                     & 9.6  & 0.8509 \\
&4.2263                   & 1056.40                    & 10.1 & 0.8489 \\
&4.2358                   & 530.30                     & 11.4 & 0.8530 \\
&4.2439                   & 538.10                     & 12.0 & 0.8552       & $22.7\pm6.0$      & $19.4\pm5.1$  & $21.5\pm5.7$                  \\
&4.2580                   & 828.40                     & 13.5 & 0.8559 \\
&4.2668                   & 531.10                     & 14.0 & 0.8587 \\
&4.2777                   & 175.70                     & 14.8 & 0.8593 \\
&4.2878                   & 502.40                     & 14.2 & 0.8612 \\
&4.3121                   & 501.20                     & 16.2 & 0.8628 \\
&4.3374                   & 505.00                     & 18.6 & 0.8646 \\
&4.3583                   & 543.90                     & 21.8 & 0.8693 \\
&4.3774                   & 522.70                     & 21.0 & 0.8668 \\ \hline
&4.3965                   & 507.80                  & 22.6 & 0.8674       &\multirow{6}*{$30.9\pm7.1$} & \multirow{6}*{$42.8\pm9.8$} &\multirow{6}*{$46.3\pm10.6$} \\
&4.4156                   & 1043.90                    & 24.8 & 0.8764 \\
&4.4362                   & 569.90                     & 25.0 & 0.8683 \\
&4.4671                   & 111.09                     & 27.4 & 0.8794 \\
&4.5271                   & 112.12                     & 30.7 & 0.8838 \\
&4.5995                   & 586.90                     & 34.0 & 0.8876 \\ \hline
&4.6152                   & 103.83                     & 33.5 & 0.8712       &\multirow{6}*{$69.4\pm11.0$} &\multirow{6}*{$54.2\pm8.6$} &\multirow{6}*{$59.0\pm9.4$} \\
&4.6304                   & 521.52                     & 34.1 & 0.8710 \\
&4.6431                   & 552.41                     & 34.5 & 0.8708 \\
&4.6639                   & 529.63                     & 35.7 & 0.8712 \\
&4.6842                   & 1669.31                    & 36.3 & 0.8711 \\
&4.7008                   & 536.45                     & 36.6 & 0.8710 \\
\hline
 \end{tabular}
\end{small}
\end{center}
\end{table*}

\section{SYSTEMATIC UNCERTAINTY}

In the cross section measurements, the systematic uncertainties mainly comes from the integrated
luminosity, tracking efficiency, PID efficiency, ISR correction, $M_{\rm rec}$ fit, and veto of
background events associated with $\Lambda$ decays.

The integrated luminosity of the data set is measured by large-angle
Bhabha scattering events, and the uncertainty in
the measurement is $1.0$\%~\cite{rlum}, which is dominated by the
precision of the MC generator used for efficiency correction.
The tracking and PID efficiencies have been studied with high purity control
samples of $J/\psi\to p\bar{p}\pi^+\pi^-$ and $\psi(3686)\to \pi^+\pi^-J/\psi\to \pi^+\pi^-p\bar{p}$ decays~\cite{prd91112004,prd99031101}.
The differences of the tracking and PID efficiencies between data and MC simulation in different
transverse momentum and total momentum ranges are obtained separately.
The averaged differences for the tracking (PID) efficiencies are corrected by the factors $f_{\rm trk}$ ($f_{\rm PID}$) as mentioned in Sec.~\ref{subsec:avgX}.
The uncertainties of the tracking and PID efficiencies are reweighted by the $p$/$\bar{p}$
and $\pi^{+}$/$\pi^{-}$ momenta of the signal MC simulation events. The reweighted uncertainties for tracking (PID)
efficiencies,
$0.1$\% ($0.3$\%) per $p$, $0.1$\% ($0.4$\%) per $\bar{p}$, $1.0$\% ($0.5$\%) per $\pi^{+}$ and
$0.8$\% ($0.4$\%) per $\pi^{-}$, are assigned as the systematic uncertainties.
Adding them linearly gives the total systematic uncertainties due to the tracking and PID efficiencies to be
$1.1$\% and $1.6$\% for the process $e^+e^-\to p p \bar{p} \bar{n} \pi^{-}$, and
$1.3$\% and $1.9$\% for the process $e^+e^-\to p \bar{p} \bar{p} n \pi^{+}$, respectively.

The input Born cross sections in the generator are iterated until the
$(1+\delta^{\gamma}) \cdot \epsilon$ values converge. The largest difference of $(1+\delta^{\gamma}) \cdot \epsilon$
between the last two iterations at all energy points, $3.2$\%, is taken as the corresponding systematic uncertainty.

Three different tests were performed to estimate the uncertainty associated with the $M_{\rm rec}$ fit.
The fit range is increased or decreased by $5$ MeV/$c^2$.
The background shape is replaced with a second-order Chebychev polynomial function, and the signal shape is replaced with an MC simulation-derived shape convolved with a Gaussian function.
The quadrature sum of these changes, $3.6$\%, is taken as the relevant uncertainty.

The systematic uncertainty due to the veto of $\Lambda$ background events is estimated by changing the $\Lambda$ veto mass window from $\pm3\sigma$ to $\pm5\sigma$, where $\sigma$ is the invariant mass resolution and the value is 3 MeV/$c^2$.
The change of the measured cross section, $0.03$\%, is assigned as the uncertainty.

Adding the above systematic uncertainties summarized in Table~\ref{tab:totalsys} in quadrature yields the total systematic uncertainties of
$5.3$\% and $5.4$\%, for the processes $e^+e^-\to p p \bar{p} \bar{n} \pi^{-}$ and $e^+e^-\to p \bar{p} \bar{p} n \pi^{+}$,
respectively. The average systematic uncertainty,
$5.35$\%, is taken as the total systematic uncertainty in the cross section measurement
for the process $e^+e^-\to p p \bar{p} \bar{n} \pi^{-}+c.c.$.

\begin{table}[H]
\caption{The relative systematic uncertainties (in \%) from
the integrated luminosity of data set ($\mathcal{L}$),
the tracking efficiency ($\rm Trk$),
the PID efficiency (PID),
the ISR correction (ISR),
the $M_{\rm rec}$ fit (Fit), and
the $\Lambda$ veto
in the cross section measurements.} \label{tab:totalsys}
\begin{center}
\begin{small}
\begin{tabular}{ccccccccc}\hline
    Mode  & $\mathcal{L}$                &  Trk   & PID   &ISR &  Fit    & $\Lambda$ veto & $\rm Total$  \\ \hline
   $p p \bar{p} \bar{n} \pi^{-}$  & $1.0$   &$1.1$  &$1.6$&     $3.2$&  $3.6$&     $0.03$&    $5.3$ \\
   $p \bar{p} \bar{p} n \pi^{+}$  & $1.0$   &$1.3$  &$1.9$&     $3.2$&  $3.6$&     $0.03$&    $5.4$ \\
\hline
\end{tabular}
\end{small}
\end{center}
\end{table}

\section{SUMMARY}

By using the data sets taken at the center-of-mass energies between $4.16$ and $4.70$ GeV,
the process $e^+e^- \to p p \bar{p} \bar{n} \pi^{-} + c.c.$ has been observed for the first time
with a statistical significance of $11.5\sigma$.
The average Born cross sections in the three energy ranges of
(4.160, 4.380), (4.400, 4.600) and (4.610, 4.700) GeV
are measured to be $(21.5\pm5.7\pm1.2)$ fb, $(46.3\pm10.6\pm2.5)$ fb and $(59.0\pm9.4\pm3.2)$ fb, respectively, where the first uncertainties are statistical and the second systematic.
The Born cross section close to threshold is larger than would be expected from five-body phase space.
The lineshape of the average Born cross sections for the process $e^+e^- \to p p \bar{p} \bar{n} \pi^{-} + c.c.$ shows similar behaviour to that of the process $e^+e^- \to 2(p\bar{p})$.
The shape of the invariant-mass spectra of $\bar{p}\bar{n}$ and $pp\pi^-$ are in good agreement with the phase-space distributions,
thereby indicating no hexaquark or di-baryon state observed with the current data sample size.

\vspace{6mm}
{\it  The BESIII collaboration thanks the staff of BEPCII and the IHEP computing center for their strong support.}
\vspace{6mm}
\end{multicols}
\clearpage

\bibliographystyle{cpc}
\bibliographystyle{references}

\clearpage
\end{CJK*}
\end{document}